\documentclass[pra,aps,groupedaddress,twocolumn,superscriptaddress,floatfix,showpacs]{revtex4}
\usepackage{epsfig}
\usepackage{mathrsfs}
\usepackage{amsmath}
\usepackage{textcmds}
\usepackage{amssymb}
\usepackage{mathptmx}
\allowdisplaybreaks[1]
\newcommand{\be}{\begin{equation}}
\newcommand{\ee}{\end{equation}}
\newcommand{\bea}{\begin{eqnarray}}
\newcommand{\eea}{\end{eqnarray}}
\newcommand{\ket}[1]{\left|#1\right\rangle}

\newcommand{\braket}[2]{\left\langle #1 | #2 \right\rangle}

\newcommand{\bc}{\begin{center}}
\newcommand{\ec}{\end{center}}

\renewcommand{\[}{\left[}
\renewcommand{\]}{\right]}
\newcommand{\forget}[1]{}

\newcommand{\re}{{\rm e}}

\newcommand{\ri}{{\rm i\,}}
\begin{document}
\title{Ultra-Stable Matter-Wave Gyroscopy with Counter-Rotating Vortex Superpositions in Bose-Einstein Condensates \\}
\author{Sulakshana Thanvanthri}
\affiliation{Hearne Institute for Theoretical Physics, Department of Physics \& Astronomy,
Louisiana State University,
Baton Rouge, Louisiana 70803-4001, USA}
\author{Kishore T. Kapale}
\email{KT-Kapale@wiu.edu}
\affiliation{Department of Physics, Western Illinois University, Macomb, Illinois, 61455-1367, USA}
\affiliation{Hearne Institute for Theoretical Physics, Department of Physics \& Astronomy,
Louisiana State University,
Baton Rouge, Louisiana 70803-4001, USA}
\author{Jonathan P. Dowling}
\affiliation{Hearne Institute for Theoretical Physics, Department of Physics \& Astronomy,
Louisiana State University,
Baton Rouge, Louisiana 70803-4001, USA}
\begin{abstract}
Matter-wave interferometers are, in principle, orders of magnitude more sensitive than their optical counterparts. Nevertheless, creation of matter-wave currents to achieve such a sensitivity is a continuing challenge. Here, we propose the use of Optical Angular Momentum (OAM) induced vortex superpositions in Bose-Einstein Condensates (BECs) as an alternative to atom interferometers for gyroscopy. The coherent superposition of two counter-rotating vortex states of a trapped condensate leads to an interference pattern that rotates by an angle proportional to the angular velocity of the rotating trap \mdash in accordance with the Sagnac effect.  We show that the rotation rate can be easily read out and that the device is highly stable. The signal-to-noise ratio and sensitivity of the scheme are also discussed.
\end{abstract}
\maketitle

Matter-wave interferometers~\cite{AtomInterferometry}, which use atomic beams in place of coherent light beams, have been investigated extensively in the past decade in the context of metrology.  The main advantage of matter-wave interferometers (employing beam of atoms with atomic mass $m$) stems from the fact that they offer sensitivities higher by a factor of about $mc^{2}/\hbar \omega \approx 10^{10}$  in comparison with optical interferometers (employing laser light of angular frequency $\omega$)~\cite{Dowling:1998}, where $c$ is the speed of light and $\hbar$ is Planck's constant divided by $2\pi$.  The matter-wave interferometers naturally require mirrors and beam splitters for atomic beams~\cite{Kasevich:2000}, which are much more complex devices compared to their optical counterparts, resulting in added complications to the interferometric setup. Thus better schemes for interferometric metrology are actively being pursued. 

In particular for gyroscopy, which deals with measurement of angular velocity of the laboratory frame of reference, the interferometer is known as Sagnac interferometer~\cite{Sagnac:1913,Scully:Book}; it employs two counter-propagating waves (light or matter) that experience a phase shift, termed as Sagnac phase shift, proportional to the area of the interferometer and the angular velocity of the lab frame. Recently, a hybrid scheme has been proposed that uses alkali atom vapor cells, acting as a slow light medium, in the path of an optical Sagnac interferometer to obtain higher sensitivity for gyroscopy~\cite{Fschauer:2006}. There is also another proposal employing entangled, parametrically down-converted photons to attain Heisenberg-limit of sensitivity for Sagnac interferometry~\cite{GSA:2007}. In this letter, for the first time, we propose the use of vortex superpositions in Bose-Einstein Condensates (BEC)~\cite{BEC} as a Sagnac interferometer. As illustrated later this scheme provides a compact and robust scheme for gyroscopy without the need of traditional interferometers employing mirrors, beam splitters and single photon or atomic state-selective detectors.

Light carrying orbital angular momentum (OAM)~\cite{PhysicsTodayOAM} allows  coherent generation of vortices in BECs~\cite{Marzlin:1997, Nandi:2004}. In fact, two of us (KTK and JPD) have taken this idea further and proposed coherent generation of arbitrary and coherent superpositions of two counter-rotating vortices in BEC~\cite{Kapale:2005}. These schemes for coherent transfer of orbital angular momentum from light to atoms have been experimentally verified as well, albeit in freely falling clouds of atoms~\cite{Phillips:2006, Bigelow:2008}. The experimentally generated vortices, which are falling under gravity, would be naturally difficult to employ for gyroscopy. With an eye on application to gyroscopy, we extended our original proposal~\cite{Kapale:2005} further and devised a robust Stimulated Raman Adiabatic Passage (STIRAP)~\cite{Shore:1992} based transfer scheme to create a superposition of arbitrary vortices in BEC~\cite{Thanvanthri:2008}. Once such a vortex superposition is created, the condensate density distribution shows an interference pattern determined by the phase difference between the amplitudes and charges of the two vortex components. The BEC trap then behaves as an interferometer and can be used to measure small changes in the phase caused by rotation of the trap or the laboratory frame of reference. 

Two important benchmarks for any gyroscope are its sensitivity and stability. The atomic-beam gyroscopes, while having high sensitivity ($10^{-9}$s$^{-1}$ Hz$^{-1/2}$), also suffer from instability due to a drift caused by fluctuations in the beam direction and thermal expansion of the beam diameter.  Since the sensitivity of the Sagnac-effect based gyroscope is directly proportional to the area of the interferometer, there is a trade-off between sensitivity and stability. We show that employing BEC vortex superposition leads to a highly stable gyroscope.


\forget{In earlier papers~\cite{Kapale:2005, Thanvanthri:2008}, we described a method to create arbitrary vortex superpositions in a BEC using OAM beams of light and a STIRAP procedure. We will summarize the results here.}

\begin{figure}[h]
\centerline{\includegraphics[scale=0.4]{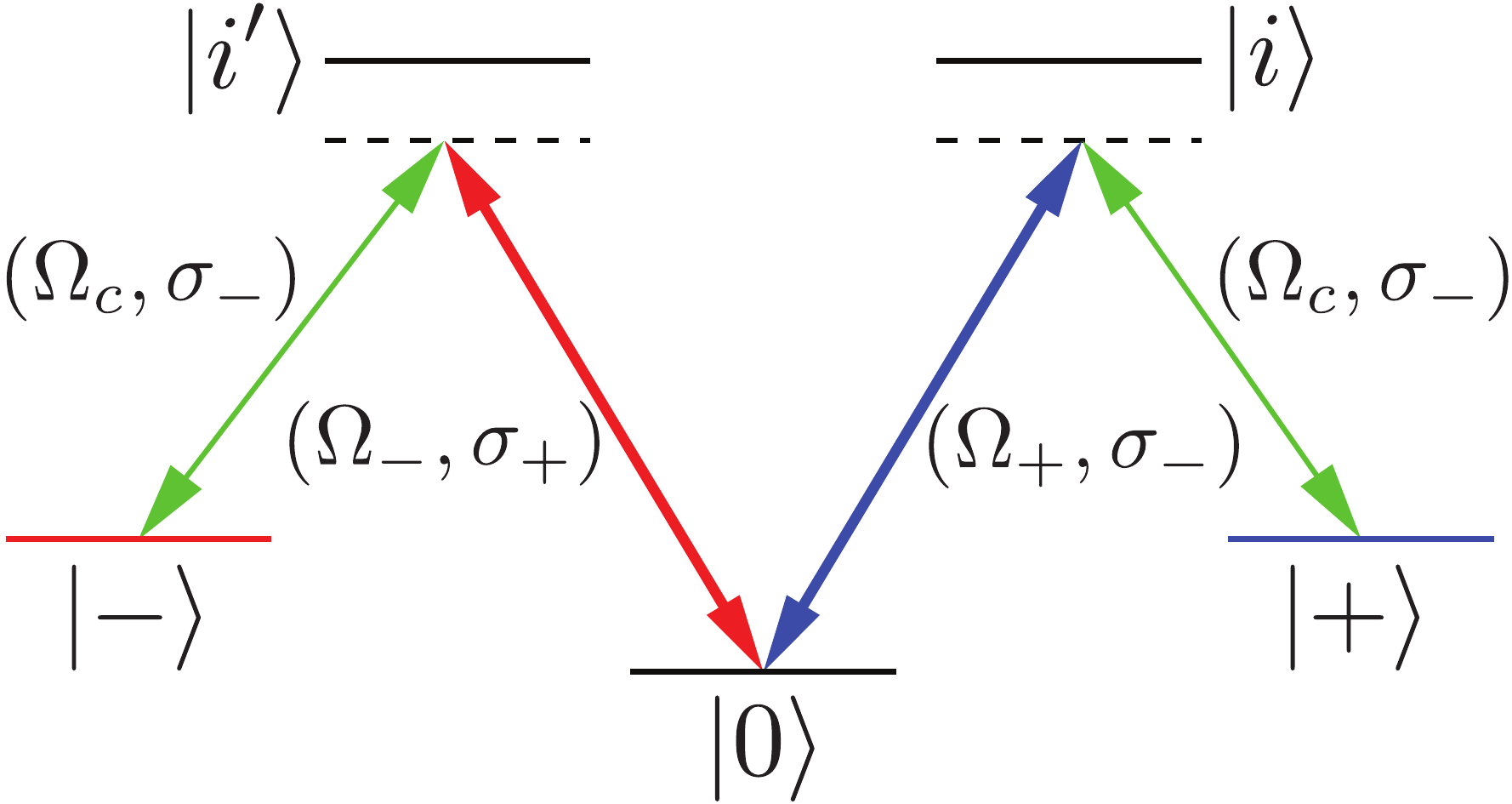}}
\caption{\label{Fig:Levelscheme} A generalized $\Lambda$ scheme that allows transfer of OAM superposition coupling dispersively with BEC atoms with Rabi frequencies $\Omega_{\pm}$ and polarization $\sigma_{+}$ via a coupling field (Rabi frequency $\Omega_{c}$ and polarization $\sigma_{-}$. The states $\ket{0}$ and $\ket{\pm}$ are only different in their rotational state and the states $\ket{i}$ and $\ket{i'}$, which are never populated, have the same rotational states as the $\ket{\pm}$ states but the hyperfine quantum numbers differ by one.}
\end{figure}

First, we briefly outline the scheme for generation of arbitrary vortex superpositions in a BEC using light carrying OAM superpositions discussed in detail in our earlier papers~\cite{Kapale:2005, Thanvanthri:2008}. The procedure involves three steps---(i) Generation of pure OAM of light by passing a normal gaussian laser beam through a computer generated hologram followed by an OAM sorter~\cite{Leach:2002}; (ii) Preparation of OAM state superposition through an interferometric scheme outlined in Ref.~\cite{Kapale:2005}; (iii) Transfer of the optical OAM superposition to a vortex superposition in BEC through a coherent Raman coupling scheme depicted in Fig.~\ref{Fig:Levelscheme}.

\begin{figure}[h]
\centerline{\includegraphics[scale=0.9]{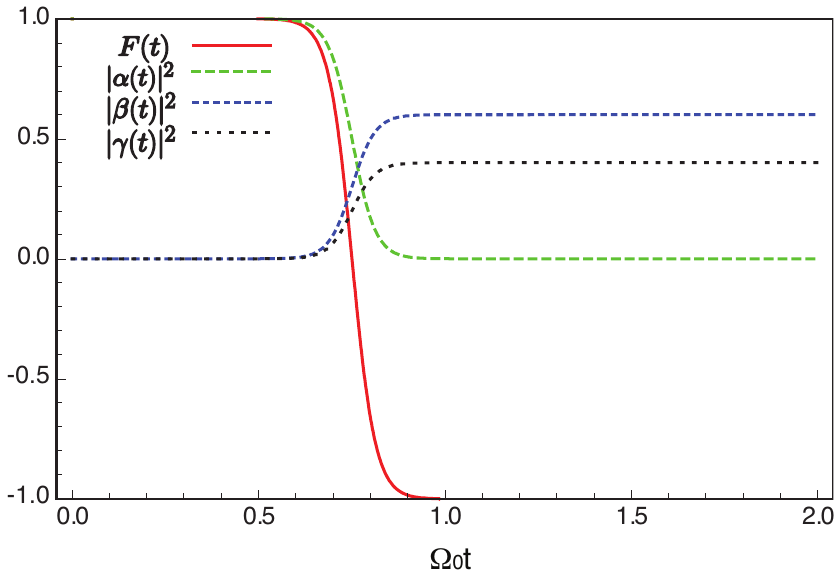}}
\caption{\label{Fig:MexHatTransfer} Generation of the vortex state superposition in a Mexican hat trap: Results of the numerical solutions of the equations for STIRAP scheme shows the superposition 60:40 of the $\ket{+}$ and $\ket{-}$ vortex states. The Rabi frequency $\Omega_{0}=1$ kHz and $\Delta = 100 ~\Omega_{0}$.}
\end{figure}

The standard three-level $\Lambda$ scheme gets unfolded into an $M$-type level scheme due to the two rotational components in the final state. The condensate is initially in the ground state $\Psi_{0}({\mathbf r}, t)$. The robust and complete transfer to an arbitrary superposition of vortex states $\Psi_{+}({\mathbf r}, t)$ and $\Psi_{-}({\mathbf r}, t)$ is possible through STIRAP like counter-intuitive pulse sequence. The spatio-temporal BEC wavefunctions, in the Thomas-Fermi (TF) approximation, are given by the ansatz:
\begin{align}
\Psi_{0}({\mathbf r}, t) &= \alpha(t) \braket{\mathbf r}{0} = \alpha(t)\, \re^{\ri (\mu/\hbar) t} \,\psi(0,\rho,\phi, z)   \\
\Psi_{+}({\mathbf r}, t) &=\beta(t) \braket{\mathbf r}{+} = \beta(t)\,\re^{\ri(\Delta+ \mu/\hbar)t} \, \psi(+l,\rho,\phi, z)\nonumber \\
\Psi_{-}({\mathbf r},t) &= \gamma(t) \braket{\mathbf r}{-} = \gamma(t)\,\re^{\ri(\Delta+ \mu/\hbar )t} \,\psi(-l,\rho,\phi, z)\,\nonumber.
\end{align}
where 
$$\psi(l,\rho,\phi, z) \propto  \frac{\rho^{|l|}e^{-il\phi}}{\sqrt{|l|!}}{\rm Re}\[\,\sqrt{\frac{\mu-V(\rho)}{\eta}}\,\,\,\] e^{-z^{2}/2}\,.$$ 
Here, $V(\rho)$ is the transverse Mexican hat atom-trap potential which is harmonic in the $z$ direction~\cite{Stringari:2006}. Also, $\mu$ is the chemical potential and the atom-atom interaction parameter $\eta = { 4 \pi \hbar a}/{m}$ where $a$ is the s-wave scattering length and $m$ is the mass of the atoms. Substituting the above wavefunctions in the Gross-Pitaevskii equations for the condensate and solving for the temporal amplitudes of different rotational states $ \alpha(t)$, $ \beta(t)$ and $\gamma(t)$ and transfer function $F(t) = |\alpha(t)|^2 - |\beta(t)|^2 - |\gamma(t)|^2 $; we find that robust and complete transfer from a non-rotating state to an arbitrary two-component vortex superposition is possible. In the long time limit $t\rightarrow \infty$,  the resulting normalized superposition state is given by:
\begin{equation}
\label{Eq:Superposition}
\Psi({\mathbf r}, t) =  b_{+} \Psi_{+}({\mathbf r}) + b_{-}  \Psi_{-}({\mathbf r})\,, 
\end{equation}
as $\alpha(t) \rightarrow 0$, $\beta(t)/\gamma(t) \rightarrow \Omega_{+}/\Omega_{-}=b_{+}/b_{-}$, where $b_{\pm}$ are the amplitudes corresponding to the two components of the OAM superposition of light coupled to the BEC. The numerical results for  the time evolution of the population of the different rotational components of the BEC and the transfer function are shown in the Fig.~\ref{Fig:MexHatTransfer}.

\begin{figure}[ht]
\includegraphics[scale=0.35]{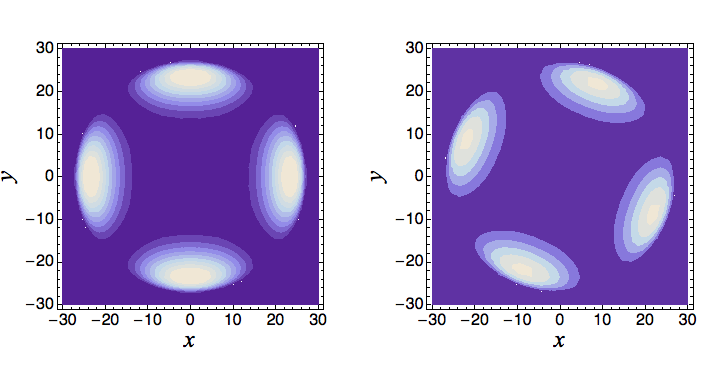}
\caption{\label{Fig:InterfPattern} Interference pattern in the atomic density produced by the OAM superposition state $(|+2\rangle+|-2\rangle)/\sqrt{2}$. The pattern is in the transverse plane for $z=0$: (a) with no rotation of the trap and (b) when angular velocity of the trap is $\Omega = 0.05$ s$^{-1}$. The lengths $x$ and $y$ are normalized with respect to the trap dimension $L_{x} = L_{y} = 2.4  \mu$m. }
\end{figure}

This superposition amounts to counter-rotating matter-wave currents causing the density profile of the BEC to be the interference pattern given by:
\begin{align}
\label{densprofile}
\Sigma({\mathbf r}, t) =  [1+{\rm cos} (2|l|\phi)] |\psi(l,\rho,z)|^2\,,
\end{align}
where $\psi(l,\rho,z)$ is the azimuthally symmetric spatial profile which is common to the two counter-rotating vortex components when they share a common vortex-charge magnitude.
This density distribution at $z=0$ is plotted in Fig. \ref{Fig:InterfPattern}. 
When the trap (or laboratory) itself rotates in time about the $z$ axis, the two counter-rotating matter waves pick up different phases and Eq.~\eqref{Eq:Superposition} is modified to $\Psi_{\Omega}({\mathbf r}, t) =  b_{+}\Psi_{+}({\mathbf r}) e^{i \phi(t)} + b_{-}\Psi_{-}({\mathbf r}) e^{-i \phi(t)}$. 

The phase $\phi(t)$ accumulated over a given time duration can be calculated using the prescription of the optical Sagnac effect, in which two optical beams counter propagate in a closed loop. In an optical interferometer with the light of wavelength $\lambda$,  the phase in one round trip of the interferometer by each component wave, is given by
$\phi_{\rm opt} ={4\mathbf{ A} \cdot \Omega}/({\lambda c})$,
where $c$ is the velocity of light, $\bf{A}$ is the area and $\Omega$, which is treated as a vector here, is the angular velocity of the interferometer~\cite{Scully:Book}.  For a BEC composed of atoms with atomic mass $m$, $\lambda c \rightarrow \hbar/m$, and the area of the interferometer is given by the cross-sectional area of the trapped BEC cloud.  Assuming that the interferometer rotates about a $z$ axis the Sagnac phase accumulated is given by~\cite{Dowling:1998}
\begin{equation}
\label{sagnacphase}
\phi_{\Omega} (t) = N(t)  \frac{4 A m}{\hbar} \Omega\,,
\end{equation}
where $N(t) $ is the number of times an atom goes around the trap center in time $t$.  
Noting that each atom in the BEC cloud has the same magnitude of the angular momentum we can determine the linear velocity, period ($T$) and as a result $N(t)= t/T$, as a function of time and $l$ to arrive at $N(t) = t \hbar l /(2 m  \pi r^2)=t \hbar l /(m A)$ and thus
\begin{equation}
\label{Eq:VortexSagnacPhase}
\phi_{\Omega} (t) = 2\, l\, \Omega\, t\,.
\end{equation}
Thus, the accumulated phase in time $t$ is proportional to the time, angular velocity of the laboratory frame and the magnitude of the vortex charge in the superposition. Even though the area dependence is not explicit in the above formula, the dependence on $l$ implicitly includes the number of round trips performed by each atom in unit time and the area of the interferometer. 
In this case, the transverse density distribution (at $z =0 $) is modified to
\begin{equation}
\label{densprofileomega}
\Sigma_{\Omega} (\rho, \phi, t) = \{1+ \mathrm {cos} [2 |l| \phi+\phi_{\Omega}(t)] \} \left |\psi(l,\rho,z)\right |^2\,, 
\end{equation}
which when compared with Eq.~\eqref{densprofile} signifies rotation of the interference pattern by an angle $\phi_{\Omega}(t)/(2 |l|)$ as the Sagnac phase accumulates over time. Following the rotation of a line drawn in the dark region of the interference pattern in the density profile allows determination of the angular velocity $\Omega$. 


The precision in measuring the angular velocity depends on the sensitivity of  the BEC density measurement. We suggest the phase contrast optical imaging of BEC clouds as a non-destructive technique that has been extensively used for  {\it in situ} imaging of trapped BECs~\cite{Andrews:1997, Andrews2:1997} for imaging the rotating BEC cloud. In this technique, images are formed by photons scattered coherently from a condensate. The phase modulation of the probe through the BEC cloud is converted to amplitude modulation by retarding the transmitted light. In the following we discuss the signal-to-noise ratio (SNR) and the sensitivity achievable in such a detection scheme. 

\begin{figure}[h]
\includegraphics[scale=0.6]{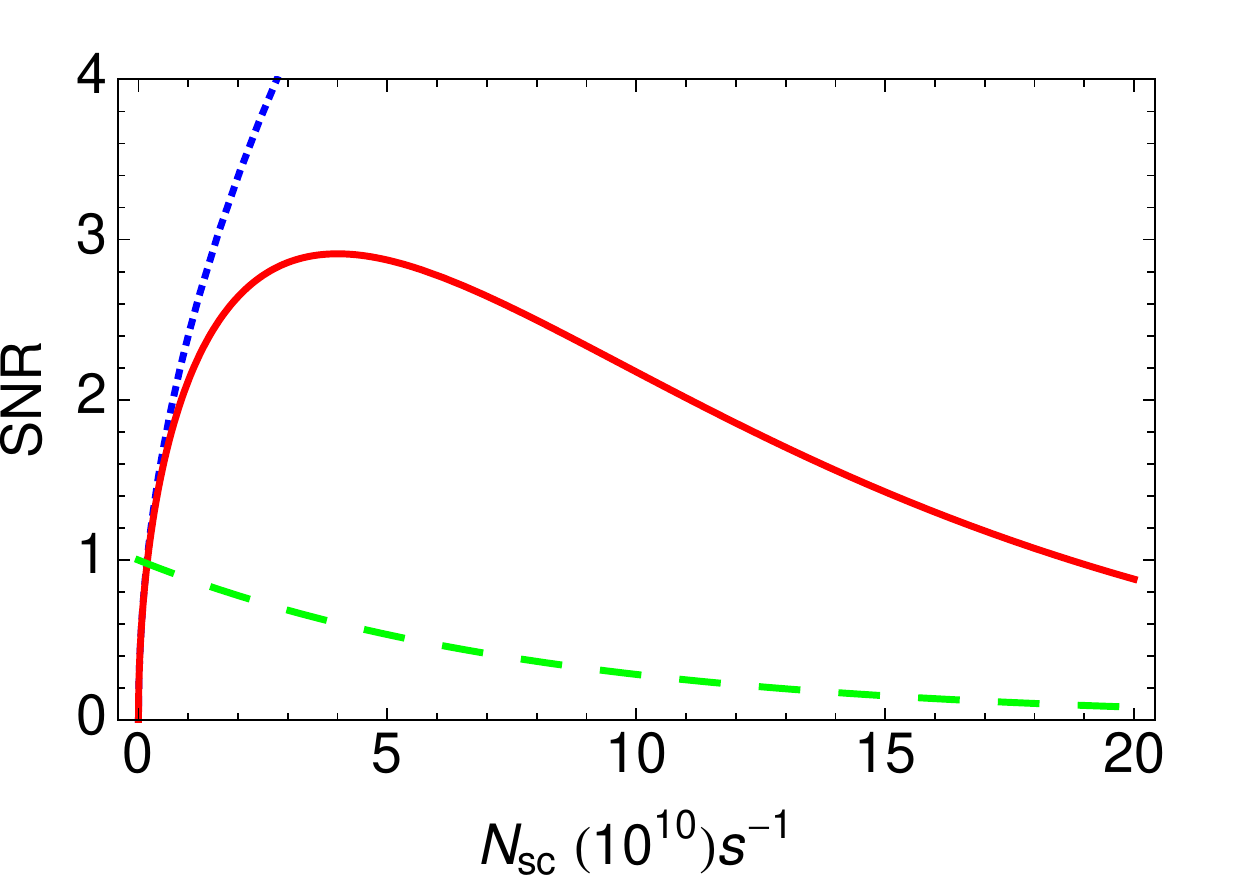}
\caption{\label{Fig:SNR}(color online) Signal-to-noise ratio (SNR) as a function of number of scattered photons arriving at the detector per second. The dotted (blue) curve gives the $1/\sqrt{N_{sc}}$ behavior, the dashed (green) curve models the exponential decay of the BEC with increase in number of probe photons. The solid (red) curve gives the effective SNR. Here the rotational velocity of the trap is taken to be $\Omega=6\times10^{-6}$ rad s$^{-1}$.}
\end{figure}

The SNR is given by the ratio of the rotational phase shift to the shot-noise that depends on the number of scattered photons $N_{\mbox{sc}}$ arriving at the detector per second~\cite{Fschauer:2006},
\begin{equation}
\label{SNR-def}
{\rm SNR} = \frac {\phi_{\Omega}}{\sqrt{1/N_{sc}}},
\end{equation}
where $N_{sc}$ is the number of scattered probe photons arriving at the detector per second from the condensate.  From Eq.~\eqref{SNR-def}, it appears that the SNR can be indefinitely increased by using higher probe-field Rabi frequency of the probe field in the phase contrast imaging scheme. However,  increasing the probe intensity enhances the risk of losing atoms from the BEC and thereby reducing the number of scattered photons from the BEC cloud and effectively reducing the SNR. An optimum balance can be reached for an atomic cloud of a given density as depicted in Fig.~\ref{Fig:SNR}.  For our current analysis, we assume that the condensate holds $10^{6}$ atoms in a pancake shaped trap with dimensions $L_{x} = L_{y} = 2.4~\mu$m and $L_{z} = 0.8~\mu$m. On  average the detector receives on the order of $10^{6}$ scattered photons per 0.1 ms i.e. $10^{10}$ photon per second from the BEC. In Fig.~\ref{Fig:SNR} we plot the SNR as a function of the number of photons detected. We model the loss of atoms from the BEC with increasing photon scattering with an exponential dependence of the SNR on the photon number. The final SNR has a maximum at a certain rate of photon detection. The sensitivity $\Omega_{{\rm{min}}}$ is calculated by setting SNR = 1.

The sensitivity of this device strongly depends on the value of the vortex charge and the general formula can be obtained by fixing SNR $ =1$ to obtain $\Omega_{{\rm min}} = 1/(2 l t \sqrt{N_{\rm sc}})$. We choose a conservative value for the magnitude of the vortex charge to be $l=100$; such a high-charge vortex can be accommodated in the Sombrero trap with appropriate choice of the trap parameters. Thus the numerical value for smallest rotational speed measurable is calculated to be $\Omega_{{\rm min}} = 5 \times 10^{-8}$ rad s$^{-1}$ Hz$^{-1/2}$ (for $t =1$ sec and $N_{\rm sc} = 10^{10}$ s$^{-1}$) which is slightly worse than the sensitivity of about $10^{-9}$ s$^{-1}$ Hz$^{-1/2}$ reported for the state-of-the-art atomic-beam gyroscopes~\cite{Arnold:2006, Gupta:2005, Fschauer:2006}. Despite the slightly lower sensitivity, our scheme offers tremendous simplicity and robustness in comparison with the other matter-wave interferometers. Furthermore, there are several avenues to enhance the sensitivity of our scheme simply by (i) improvement in the non-destructive imaging of BEC, (ii) allowing phase accumulation for longer times if accuracy is more important than speed or in the case when the rotational speed to be detected is not changing rapidly with time, and (iii) by performing simultaneous measurements on about 10000 BEC clouds arranged in a 2D optical lattice to easily surpass the current atomic-beam based gyroscopes. 

To address the stability of our proposed gyroscope, it helps to note that beam drift is an important source of instability in the long-integration-time gyroscopy employing traditional atomic  or optical beams. In BEC vortices, angular momentum is quantized and all the atoms move with a common angular velocity and besides having to maintain the trapping potential no complicated maneuvering of the atomic beams is required. Also thermal expansion over long beam trajectories is not an issue for a trapped BEC.   Thus our proposal eliminates the drift problem commonly faced by current atomic gyroscopes with drifts are on the order of $68~\mu$deg/h after about four hours of operation~\cite{Kasevich:2006}. 

The area of the interferometer is one of the important parameter governing the sensitivity; in our scheme the size of the BEC trap, which a few mm$^{2}$, is the area of the interferometer. On the other hand, the typical area of an atom beam gyroscope is about 4 m$^{2}$~\cite{Kasevich:2006}. Thus despite such a large difference in the area, current numbers for sensitivity for the vortex-based gyroscope  are only slightly worse than that of the state-of-the-art atomic gyroscopes. Several methods are already available for improving the sensitivity  of the vortex based gyroscope such as using higher density BEC clouds, using optical lattice of BECs~\cite{Kasevich:2007} and improving the sensitivity and speed of the BEC imaging techniques.  An important point to note is that for the BEC-vortex gyroscope an increase in sensitivity (by increasing the area) does not correspond to a decrease in stability. The stability can be maintained and enhanced by manipulating the geometry of the BEC trap and optical beams used to create the vortex superposition. Further, if it is necessary to replenish the trap with a fresh BEC cloud it can be noted that creating and reloading BECs has become a routine procedure with improvements in the experimental techniques coming regulary.
 

In summary, we have presented a new and simplified scheme for a matter-wave gyroscope. The superpositions of oppositely `charged' OAM beams of light are employed to coherently  transfer a non-rotating ground state BEC into a superposition of two counter-rotating vortices. The rotational momenta acquired by the BEC wave function gives an interferometric  Sagnac effect that can be used to sense the rotation of the trap itself. The slightly worse sensitivity of the proposed gyroscope compared with the corresponding experimental value of sensitivity for atomic-beam based gyroscopes is not a significant drawback in light of the advantages offered by it. The advantages include---a compact and robust setup, no need of the guiding mechanism (beamsplitters and mirrors) for atoms and elimination of the drift problem.  The sensitivity can surely be improved with improved experimental techniques.

The authors would like to thank F. Narducci and D. Sheehy for useful discussions. This work was supported by the Army Research Office, the Defense Advanced Research Projects Agency, and the Intelligence Advanced Research Projects Agency.


\begin{thebibliography}{10}

\bibitem{AtomInterferometry} P.~R. Berman, {\em Atom Interferometry}  (Academic Press, San Diego, 1996)

\bibitem{Dowling:1998} J.~P. Dowling, Phys. Rev. A {\bf 57},  4736  (1998). 

\bibitem{Sagnac:1913} 
G. Sagnac, Comptes Rendus {\bf 157} 708--710 (1913), english translation available at: http://zelmanov.ptep-online.com/papers/zj-2008-07.pdf; G. Sagnac, Comptes Rendus {\bf 157} 1410--1413 (1913), english translation available at: http://zelmanov.ptep-online.com/papers/zj-2008-08.pdf.

\bibitem{Scully:Book} M. O. Scully and M. S. Zubairy {\em Quantum Optics} (Cambridge University Press, Cambridge, 1997).

\bibitem{Kasevich:2000} T. L. Gustavson, A. Landragin and M. A. Kasevich, Class. Quantum Grav. {\bf 17}, 2385 (2000).

\bibitem{Fschauer:2006} F. E. Zimmer and M. Fleischhauer, Phys. Rev. A {\bf74}, 063609  (2006). 

\bibitem{GSA:2007} A. Kolkiran and G. S. Agarwal, Opt. Exp. {\bf 15}, 6798--6808 (2007).

\bibitem{BEC}
M.~H. Anderson {\it et~al.}, Science {\bf 269},  198  (1995);
C.~C. Bradley, C.~A. Sackett, J.~J. Tollett, and R.~G. Hulet, Phys. Rev. Lett.
  {\bf 75},  1687  (1995);
M.-O. Mewes {\it et~al.}, Phys. Rev. Lett. {\bf 77},  416  (1996).

\bibitem{PhysicsTodayOAM} M. Padgett, J. Courtial and L. Allen, Phys. Today {\bf 57}, 35--40 (May 2004).

\bibitem{Marzlin:1997}
K.-P. Marzlin, W. Zhang, and E.~M. Wright, Phys. Rev. Lett. {\bf 79},  4728
  (1997).

\bibitem{Nandi:2004}
G. Nandi, R. Walser, and W.~P. Schleich, Phys. Rev. A {\bf 69},  063606
  (2004).

\bibitem{Kapale:2005}
K.~T. Kapale and J.~P. Dowling, Phys. Rev. Lett. {\bf 95},  173601  (2005).

\bibitem{Phillips:2006}
K. Helmerson et. al., Proc. of SPIE, {\bf 6326}, 632603 (2006).

\bibitem{Bigelow:2008}
K. C. Wright, L. S. Leslie and N. P. Bigelow, Phys. Rev. A {\bf 77},  041601  (2008). 

\bibitem{Shore:1992}
B.~W. Shore {\it et~al.}, Phys. Rev. A {\bf 45},  5297  (1992).

\bibitem{Thanvanthri:2008}
S. Thanvanthri, K.T. Kapale and J. P. Dowling, Phys. Rev. A {\bf  77}, 053825 (2008).

\bibitem{Leach:2002}
J. Leach {\it et~al.}, Phys. Rev. Lett. {\bf 88},  257901  (2002).

\bibitem{Stringari:2006}
M. Cozzini, B. Jackson and S. Stringari, Phys. Rev. A {\bf 73}, 013603 (2006).

\bibitem{Andrews:1997}
M. R. Andrews {\it et ~al.}, Science {\bf 275},  637 (1997).

\bibitem{Andrews2:1997}
M. R. Andrews {\it et ~al.}, Phys. Rev. Lett., {\bf 79}, 553 (1997).

\bibitem{Arnold:2006}  A. S. Arnold, C. S. Garvie, and E. Riis, Phys. Rev. A {\bf 73},
041606 R (2006).

\bibitem{Gupta:2005} S. Gupta, K. W. Murch, K. L. Moore, T. P. Purdy, and D. M.
Stamper-Kurn, Phys. Rev. Lett. {\bf 95}, 143201 (2005).

\bibitem{Kasevich:2006} D. S. Durfee, Y. K. Shaham, and M. A. Kasevich, Phys. Rev. Lett. {\bf 97}, 240801 (2006).

\bibitem{Kasevich:2007} W. Li, A. K. Tuchman, H. Chien, and M. A. Kasevich, Phys. Rev. Lett. {\bf 98}, 040402 (2007).



















%












%






\end{thebibliography}
\end{document}